\documentclass[twocolumn,showpacs,prb,preprintnumbers,amsmath,amssymb]{revtex4}

\usepackage{graphicx}
\usepackage{dcolumn}
\usepackage{bm}

\newcommand{\bol}[1]{\boldsymbol #1}

\begin{document}

\title{Coupled S $\bol =$ $\bol 1 \over\bol 2$ Heisenberg antiferromagnetic chains 
in an effective staggered field}

\author{Masahiro Sato}
\author{Masaki Oshikawa}
\affiliation{Department of Physics, Tokyo Institute of Technology, 
Oh-okayama, Meguro-ku, Tokyo 152-8550, Japan}

\begin{abstract}
We present a systematic study of coupled $S=1/2$ Heisenberg antiferromagnetic
chains in an effective staggered field. We investigate several effects of 
the staggered field in the {\em higher} ({\em two or three}) 
{\em dimensional} spin system analytically. 
In particular, in the case where the staggered field and the
inter-chain interaction compete with each other, 
we predict, using mean-field theory, a characteristic phase transition. 
The spin-wave theory predicts that the behavior of the gaps induced 
by the staggered field is different between the competitive case and the 
non-competitive case. When the inter-chain interactions
are sufficiently weak,
we can improve the mean-field phase diagram by using chain mean-field
theory and the analytical results of field theories.  
The ordered phase region predicted by the chain mean-field theory is
substantially smaller than that by the mean-field theory.
\end{abstract}

\maketitle

\section{Introduction}
\label{Sec1}
The effects induced by magnetic fields in magnets have been a 
subject of theoretical research interest for a long time. In particular, 
recently the magnetization processes of various spin chains and ladders 
have been investigated intensively. Owing to the progress of the
various experimental methods, there has been an increasing connection
between the theories and the experiments. In such a context, one of the
new attractive subjects in magnetism is the effects of a {\em staggered}
magnetic field, namely, a magnetic field which changes direction alternatingly.
While it may sound unrealistic,
there exist at least three mechanisms
generating the staggered fields in real magnets, as discussed 
in Refs.~\onlinecite{O-A} and \onlinecite{Ladd-Wang}. 

The first mechanism is due to the staggered gyromagnetic ($g$) tensor,
which can be present if the crystal structure is
not translationally invariant.
The staggered $g$ tensor $g_{\alpha \beta}^{st}$ is
defined in the coupling between the spin and the external magnetic
field (Zeeman term) as
\begin{equation}
\label{eq1-3}
\hat H_{\rm Zeeman}  = - \mu_{B} \sum_{j}H_{\alpha}
[g_{\alpha \beta}^{u}+(-1)^{j}g_{\alpha \beta}^{st}]  
S^{\beta}_{j},
\end{equation}
where $\vec H=(H_{x},H_{y},H_{z})$ is an applied uniform magnetic field
and $S^{\beta}_{j}$ is the spin operator of the local magnetic moment.
Here $H_{\alpha}g_{\alpha
\beta}^{st}$ is nothing but an effective staggered field.
In addition, the staggered field may also arise from 
the staggered Dzyaloshinskii-Moriya (DM) interaction \cite{Dzyal,MoriyaDM}
\begin{equation}
\label{eq1-2}
\hat H_{\rm DM} =  \sum_{j} (-1)^{j} \vec D \cdot
\left(\vec S_{j} \times \vec S_{j+1} \right),
\end{equation}
which can be present if the crystal symmetry is sufficiently low.
It is shown in Refs.~\onlinecite{O-A}and \onlinecite{A-O} that in the presence of the
staggered DM interaction along the chain, an applied uniform field 
$\vec H$ also generates an effective staggered field 
$\vec h\propto\vec D\times \vec H$. Several quasi-one-dimensional
Heisenberg antiferromagnets are now known to have the staggered field
due to the above mechanisms. The well known examples are   
Cu-benzoate,~\cite{Cu-Den,Cu-Aji1,Cu-Aji2}
$\rm [PM\cdot Cu(NO_{3})_{2}\cdot (H_{2}O)_{2}]_{n}$ 
(PM$=$pyrimidine),~\cite{PMCu-ex} and 
$\rm Yb_{4}As_{3}$.~\cite{YbAs-Oshi,YbAs-Shiba,Yb4As3-3} 
All of these have low-symmetry crystal structures which allow a
staggered $g$ tensor and a DM interaction along the chain.~\cite{Note1}
It is expected in Refs.~\onlinecite{Alcaraz,O-A} and \onlinecite{A-O} 
that the staggered field induces an excitation gap in the $S=1/2$ Heisenberg
antiferromagnetic (HAF) chain, which should be otherwise gapless.
The excitation gap caused by the staggered field is indeed 
found in these materials.~\cite{Cu-Den,PMCu-ex,Yb4As3-3}
Moreover, low-temperature anomalies in physical quantities, such as 
the susceptibility and the electron spin resonance line-width,~\cite{ESR-OA-PRL,ESR-OA}
are also successfully explained as effects of the staggered
field~\cite{A-O}. 
Thus it is confirmed that they are described by an 
$S=1/2$ HAF model with an effective staggered field.

There is another, rather different,
mechanism to generate a staggered field. Let us suppose that
the system consists of two sublattices, with
a weak inter-lattice coupling and strong intra-lattice one.  
If one of the sublattices is N\'eel ordered, 
the inter-lattice coupling, as a mean field,
could give an effective staggered field on the other sublattice.
The realization of this scenario is in $\rm R_{2}BaNiO_{3}$ 
where $\rm R$ is a magnetic rare earth, and the $\rm R$-ion lattice 
provides a staggered field
for $\rm Ni$ chains ($S=1$).~\cite{R2BaNi-PRL,R2BaNi-PRB}

Actually all the materials discussed above are highly one-dimensional (1D).
However, at lower temperature and lower energy, the inter-chain
interaction will eventually be dominant. In addition, there are reports 
on a few materials [$\rm CuCl_{2}\cdot 2DMSO$
(DMSO$=$dimethylsulphoxide) (Refs.~\onlinecite{2DMSO-Ex,2DMSO-Dan,2DMSO-Kenz}) and 
$\rm BaCu_{2}(Si_{1-x}Ge_{x})_{2}O_{7}$ (Refs.~\onlinecite{BaCu2-x})] 
which seem to have an effective staggered field and also a relatively 
large inter-chain interaction. 
Therefore the work including the inter-chain interaction
could be relevant for experiments. 
 
Given these backgrounds, in the present paper, we would like to clarify
the characteristic roles of staggered fields in
{\em higher dimensional} spin systems.
In this paper, we are concerned with dimensions {\em higher than 1} but
still realistic in condensed matter physics,
namely, {\em two or three} dimensions.
However, most of the analyses in this paper apply straightforwardly
even to four dimensions or higher.

Varieties of spin models with effective staggered fields are conceivable.
As a simplest model including
both the staggered field and the inter-chain coupling,
in this paper, we concentrate on the following $S=1/2$ spatially 
anisotropic Heisenberg Hamiltonian
\begin{eqnarray}
\label{eq1-1}
\hat H & = & \sum_{\vec r}\Big(J\vec S_{i,j,k} \cdot \vec S_{i+1,j,k} 
+ J_{\perp}\vec S_{i,j,k} \cdot \vec S_{i,j+1,k} \\ \nonumber
& & +J_{\perp}'\vec S_{i,j,k}\cdot \vec S_{i,j,k+1}\Big)\\ \nonumber
& &-H  \sum_{\vec r}S_{i,j,k}^{z}-h \sum_{\vec r}(-1)^{i}S_{i,j,k}^{x},
\end{eqnarray}
where $\vec S_{i,j,k}$ is the spin $1/2$ operator on $\vec r=(i,j,k)$ site. 
The coupling constants are restricted to $J > |J_{\perp}|\geq |J_{\perp}'|$,  
and thus the $i$ direction is the strongly antiferromagnetic (AF) coupled one. 
The system with $J_{\perp}'=0$ is 2D, in which the index $k$ vanishes. 
The last two terms represent the uniform and staggered
Zeeman terms respectively. 

In our model (\ref{eq1-1}), one can immediately find that 
when the inter-chain interactions are AF, they compete with the staggered Zeeman energy,
while in the ferromagnetic (FM) case, both the interactions and the
staggered field $h$ jointly make a N\'eel state stable.
Let us refer the AF case as the {\em competitive} case, and the FM case as 
the {\em non-competitive} one.
As will be explained later on, we predict 
that the competition brings a second-order phase transition in the
competitive case. 
Its emergence is one of the most characteristic
effects of the staggered field in our higher dimensional spin model.

The rest of this paper is organized as follows.
In Sec.~\ref{Sec2}, we apply mean-field theory to the model (\ref{eq1-1}). 
In the competitive case a phase transition is predicted. 
Since we are primarily interested in the transition, which is
characteristic for the higher dimensional system, 
in the later sections we will mainly discuss the competitive case. 
The non-competitive case is touched as a comparison to the competitive
case. Besides the phase diagrams, 
the mean-field magnetization curves and critical exponents are derived
from the self-consistent equations. 
In Sec.~\ref{Sec3}, using linear spin-wave approximation, 
we derive the spin-wave dispersions in the
the competitive case and in the non-competitive case. 
As a result, we find that the excitation gap
induced by the staggered field behaves differently
between the competitive case and the non-competitive case. 

In Sec.~\ref{Sec4}, we improve the mean-field phase diagrams 
by using chain mean-field theory.~\cite{CMF-Scal,CMF-Sch}
The latter is 
expected to be superior, when the inter-chain interactions are
weak and the effective 1D model can be solved exactly.
The improved diagram shows that in the weakly coupling region of the
competitive case, the ordered phase becomes much narrower than the
mean-field prediction. In the last section, we summarize those results and discuss 
future problems. In Appendix, the details of the spin-wave results are given.

\section{Mean-Field Theory Approach}
\label{Sec2}
In this section, we treat the model (\ref{eq1-1}) within 
mean-field theory (MFT) framework. We first discuss the competitive
case, and then touch the non-competitive case briefly.
\begin{figure}
\scalebox{0.8}{\includegraphics{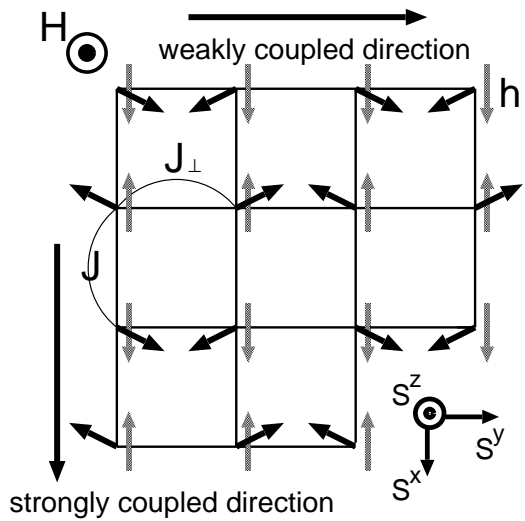}}
\caption{Directions of the magnetic fields and the spin moments 
in the 2D competitive case. The short black arrows are the spin moments
 projected onto spin $xy$ plane in the ordered phase (spontaneous symmetry
 breaking phase, see text) expected by
 the MFT. The gray arrows indicate the direction of the staggered field
 $h$. The uniform field $H$ is applied perpendicular to this paper.}
\label{Fig2}
\end{figure}

In the {\rm competitive} case, considering the advantage of both the
inter-chain energy and the staggered Zeeman energy as well as the
intra-chain coupling $J$, we can expect 
that the spin moment turns as Fig.~\ref{Fig2} at sufficiently low
temperature and small fields comparable to
the inter-chain couplings. The assumed spin moment is 
\begin{eqnarray}
\label{eq2-1}
\langle \vec S_{i,j,k} \rangle_{\rm MFT} =  \left(
\begin{array}{lll}
(-1)^{i}m_{x}, & (-1)^{i+j+k}m_{y}, & m_{z} 
\end{array}
\right). 
\end{eqnarray}
The choice of the mean field (\ref{eq2-1}) is presumably valid as long as $J$ is
sufficiently larger than $J_{\perp}$ and $J_{\perp}'$. 
On the other hand, for instance, in the limiting case: 
$J\rightarrow 0$, where the model is reduced to
a 2D or 1D AF one with a uniform field, there are possibilities that 
$\langle S_{j,k}^{x,z}\rangle$ are inhomogeneous along the $j$ or $k$ directions, 
and thus Eq. (\ref{eq2-1}) is invalid. In this paper, the MFT 
in the competitive case is performed only within the mean field (\ref{eq2-1}).

\begin{figure}[h]
\scalebox{0.7}{\includegraphics{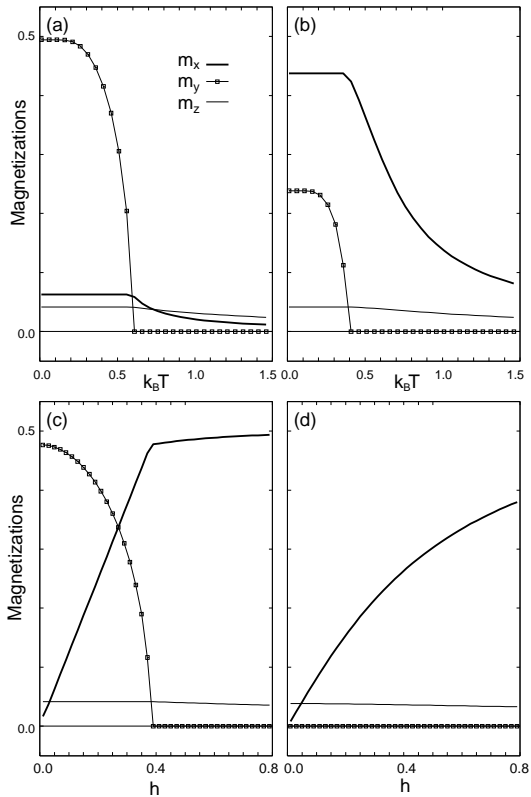}}
\caption{Magnetization curves of the competitive case for 
$(J,J_{\perp}+J_{\perp}',H)=(1,0.2,0.2)$. These are obtained by 
the numerical iterative method for Eq. (\ref{eq2-2-1}).  
The upper two parts (a) and (b) are in $h=0.05$ and $h=0.35$, 
respectively. The lower two parts (c) and (d) are in $k_{B}T=0.3$ 
and $k_{B}T=0.7$, respectively.}
\label{Fig3}
\end{figure}

\begin{figure}[h]
\scalebox{0.5}{\includegraphics{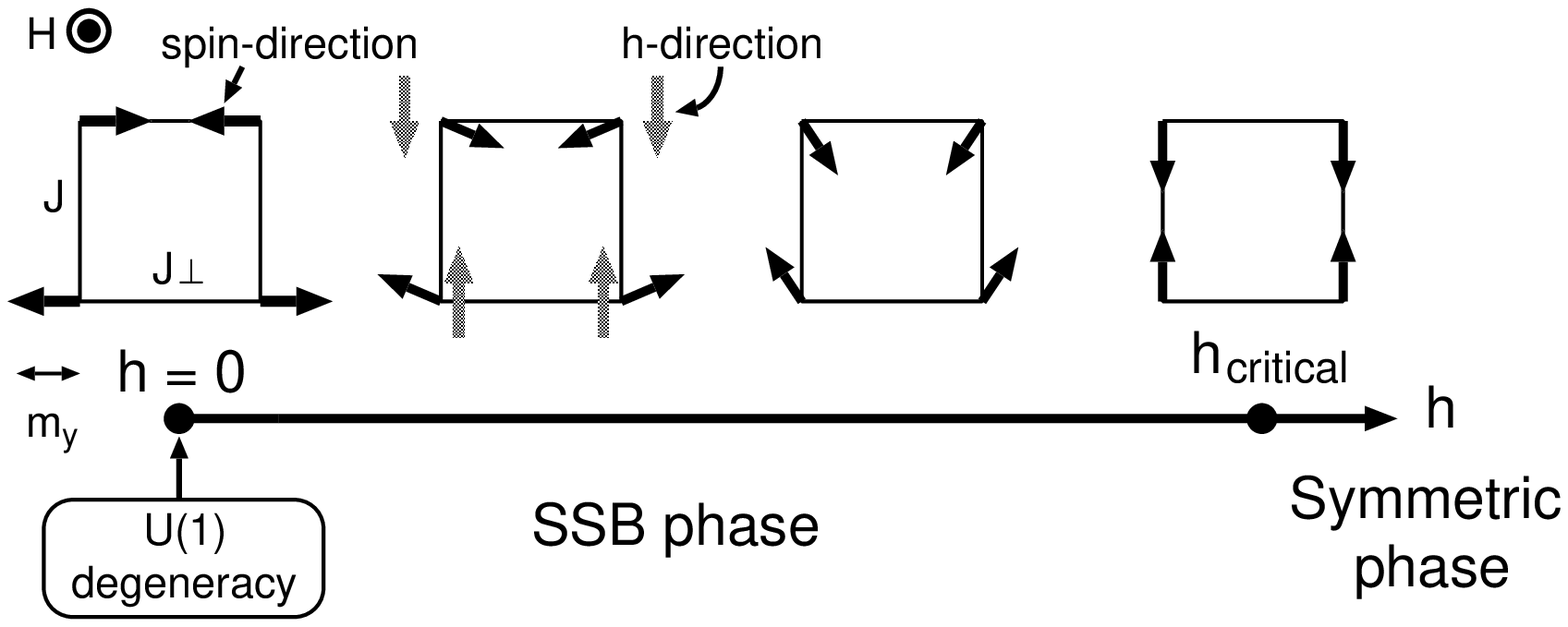}}
\caption{Variation of the spin configuration in a plaquette 
when the staggered field is increased gradually in the 2D competitive
 case. The black arrows represent spin directions projected onto spin 
$xy$ plane. The gray arrows are the staggered field $h$.}
\label{Fig4}
\end{figure}

The minimal condition for the mean-field free energy gives the
following self-consistent equations:
\begin{subequations}
\label{eq2-2ful}
\begin{eqnarray}
m_{\alpha} &=& \frac{\epsilon_{\alpha}}{2\epsilon} \tanh(\beta \epsilon),\\ 
\label{eq2-2-1} 
m^{2} &=&  \frac{1}{4}\tanh^{2}(\beta \epsilon),\label{eq2-2}
\end{eqnarray}
\end{subequations}
where
\begin{eqnarray}
\label{eq2-3}
\left\{
\begin{array}{lll}
\epsilon_{x}  &\equiv& (J-J_{\perp}-J_{\perp}^{\prime})m_{x}+h/2 \\
\epsilon_{y}  &\equiv& (J+J_{\perp}+J_{\perp}^{\prime})m_{y} \\
\epsilon_{z}  &\equiv& -(J+J_{\perp}+J_{\perp}^{\prime})m_{z}+H/2 \\
\epsilon      &\equiv& (\epsilon_{x}^{2}+\epsilon_{y}^{2}
+\epsilon_{z}^{2})^{1/2}
\end{array}
\right.,
\end{eqnarray}
and $\beta$ and $m$ are, respectively, the inverse temperature
$\frac{1}{k_{B}T}$ and the total magnetization per site. 
The numerical solutions of Eq. (\ref{eq2-2-1}) are given in Fig.~\ref{Fig3}. 
They indicate that there is a second-order phase transition, 
and the corresponding order parameter is the $y$ component
of the spin moment $m_{y}=|\langle S_{i,j,k}^{y} \rangle |$.  
Going back to Fig.~\ref{Fig2}, one sees that the phase with finite
$m_{y}$ breaks the translational symmetry in the weakly coupled
direction. 
In the following, we call this phase as the {\em SSB} 
(spontaneous symmetry breaking) phase. The other phase, in which the
spins are aligned to the field with $m_{y}=0$, will be called as the
{\em symmetric} phase. We emphasize that the transition between these two phases 
occurs only in the high dimensional spin systems and in the presence
of the competition.  
From these results, we can illustrate the variation of 
the spin moment when the staggered field $h$ is increased gradually 
at small $T$ and $H$ with Fig.~\ref{Fig4}.

\begin{figure}
\scalebox{0.6}{\includegraphics{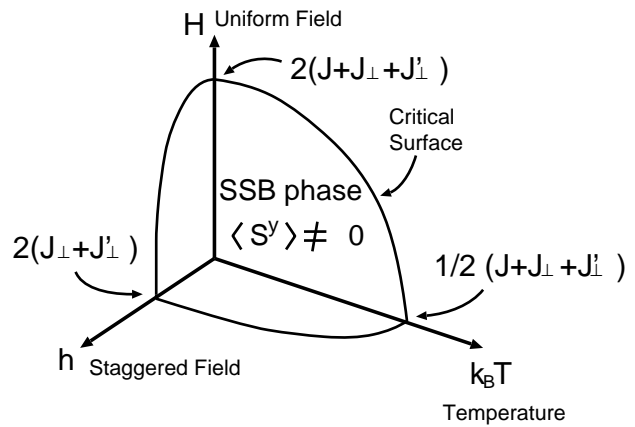}}
\caption{Schematic mean-field phase diagram in the competitive case.}
\label{Fig5}
\end{figure}

In the SSB phase, the relations
$m_{x}= \frac{h}{4(J_{\perp}+J_{\perp}')}$ and 
$m_{z}= \frac{H}{4(J+J_{\perp}+J_{\perp}')}$ hold within the
MFT. Inserting these into Eq.~(\ref{eq2-2}) and 
taking the limit $m_{y}\rightarrow 0$, we obtain the mean-field
critical surface in the space $(k_{B}T,H,h)$: 
\begin{eqnarray}
\label{eq2-4}
\tilde h_{c}^{2}+\tilde H_{c}^{2}
=\frac{1}{4}\tanh^{2} \left\{ \beta_{c}(J+J_{\perp}+J_{\perp}')
\sqrt{\tilde h_{c}^{2}+\tilde H_{c}^{2}}\right\},
\end{eqnarray}
where $\tilde h_{c}\equiv\frac{h_{c}}{4(J_{\perp}+J_{\perp}')}$, 
$\tilde H_{c}\equiv\frac{H_{c}}{4(J+J_{\perp}+J_{\perp}')}$ 
and the subscript $c$ represents critical values. 
It can be simplified in the cases $T=0$, $h=0$ and $H=0$, respectively, as
\begin{eqnarray}
\label{eq2-5}
\left\{
\begin{array}{cc}
\tilde h_{c}^{2}+\tilde H_{c}^{2}= \frac{1}{4} &  \\
\tilde H_{c}=\frac{1}{2} \tanh\left(\frac{\beta_{c} H_{c}}{4}\right) &  \\
\tilde h_{c}= \frac{1}{2} 
\tanh\left\{\beta_{c}(J+J_{\perp}+J_{\perp}')\tilde h_{c} \right\} &
\end{array}
\right..
\end{eqnarray}
Thus the mean-field phase diagram can be represented as Fig.~\ref{Fig5}.
Using the critical condition, one can calculate some critical exponents within the MFT. 
Near the critical surface in the SSB side, the order parameter,
the off-diagonal uniform and staggered susceptibilities: 
$\chi_{u}\equiv \frac{\partial m_{y}}{\partial H}$ 
and $\chi_{s}\equiv \frac{\partial m_{y}}{\partial h}$ behave
, respectively, as $m_{y}\sim (A_{c}-A)^{\beta}$, 
$\chi_{u}\sim -(A_{c}-A)^{\gamma}$ and $\chi_{s}\sim -(A_{c}-A)^{\gamma'}$
where $A$ stands for $T$, $H$ or $h$. The critical exponent
$\beta$ is found to be the conventional mean-field value $1/2$.
On the other hand, both $\gamma$ and $\gamma'$ turns out to be $1/2$,
which is different from the standard MFT result $1$.
This is because $m_{y}$ is perpendicular to $H$ and $h$, and thus the
latter are not the conjugate field as in the standard case. 
The mean-field energy per site is given as 
\begin{eqnarray}
\label{eqA-5}
\epsilon_{\rm MFT} &=& - \Big\{J(m_{x}^{2}+m_{y}^{2}-m_{z}^{2})
+Hm_{z}+hm_{x} 
\nonumber \\
&& +(J_{\perp}+J_{\perp}')(m_{y}^{2}-m_{x}^{2}-m_{z}^{2}) \Big\},
\label{eq:MFTenergy}
\end{eqnarray}
From this, one can easily confirm that the critical exponent
for the specific heat is zero. 

\begin{figure}
\scalebox{0.5}{\includegraphics{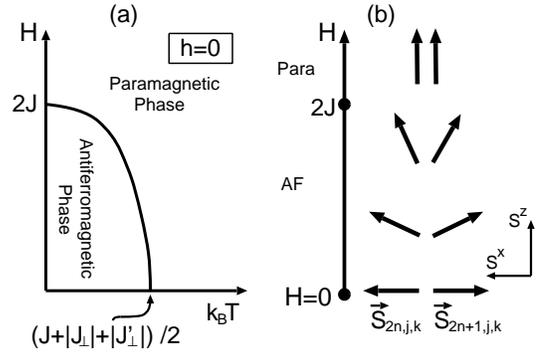}}
\caption{(a) Schematic mean-field phase diagram of the non-competitive case
 in $h=0$. (b) the variation of the spin moment when the uniform field is
 increased gradually in $h=T=0$.}
\label{Fig6}
\end{figure}

Now, we turn to the non-competitive case. 
Because of no competitions, no singular phenomena occur when $h\neq 0$. 
Canting of the spins in $zx$ plane lowers both the inter-chain
interactions and the Zeeman energies.
Thus the expectation value of the spin moments can be put as
\begin{eqnarray}
\label{eq2-7}
\langle \vec S_{i,j,k} \rangle_{\rm MFT} 
& = & \left(
\begin{array}{lll}
(-1)^{i}m_{x}, & 0, & m_{z}
\end{array}
\right).
\end{eqnarray} 
The MFT in this case gives the self-consistent equations
$m_{x(z)}= (\epsilon_{x(z)}'/2\epsilon') \tanh(\beta \epsilon')$
where $\epsilon_{x(z)}'\equiv
[(-)J+|J_{\perp}|+|J_{\perp}'|]m_{x(z)}+h(H)/2$
and $\epsilon'\equiv (\epsilon_{x}^{2}+\epsilon_{z}^{2})^{1/2}$. 
At $h=0$, the system reduces to a conventional AF magnet in a uniform
magnetic field. 
Hence, there must be a phase transition which divides the AF and
paramagnetic phases, characterized by the order parameter $m_{x}$. 
The critical line is given by
\begin{equation}
\label{eq2-8}
\frac{H_c}{4J} = \frac{1}{2} \tanh 
\left\{\beta_c (J+|J_{\perp}|+|J_{\perp}'|) \frac{H_c}{4J} \right\}.
\end{equation} 
The phase diagram and the variation of the spin moment 
are drawn in Fig.~\ref{Fig6}. 
In the three-dimensional parameter space $(k_{B}T,H,h)$,
the AF phase gives a first-order phase transition plane.

\section{Linear Spin-Wave Approximation in $T=0$}
\label{Sec3}
With the MFT described in the preceding section, we investigate 
the effects of the quantum fluctuations in both the competitive and 
the non-competitive cases, at $T=0$. The standard linear spin-wave 
approximation, based on the Holstein-Primakoff transformation 
(HPT) (Ref.~\onlinecite{HPtrans}) is employed.
The detailed results are given in Appendix \ref{App2}.

First we discuss the SSB phase of the competitive case, which is the
main subject.
In the HPT, we replace the spin operator 
with a boson annihilation (creation) operator $c$ ($c^{\dag}$) as follows:
\begin{equation}
\label{eq3-1}
\vec S_{\rm HP} 
= \left(
\begin{array}{c}
i\sqrt{\frac{S}{2}(1-\frac{c^{\dag}c}{2S})}(c^{\dag}-c) \\
S-c^{\dagger}c \\
\sqrt{\frac{S}{2}(1-\frac{c^{\dag}c}{2S})}(c^{\dag}+c)
\end{array}
\right)
\approx \left(
\begin{array}{c}
i\sqrt{\frac{S}{2}}(c^{\dag}-c) \\
S-c^{\dagger}c \\
\sqrt{\frac{S}{2}}(c^{\dag}+c)
\end{array}
\right),
\end{equation}
where $S$ is the spin quantum number generalized from $S=1/2$ and the 
sign $\approx$ denotes the leading approximation in the $1/S$ expansion. 
The above HPT is useful if the spin points to the $y$ direction
in the classical ground state, as
the bosons then represent quantum fluctuations.

In the SSB phase, actually, a canting
structure~(\ref{eq2-1}) is expected in the classical ground state,
which is equivalent to the MFT at $T=0$.
The canted spin moments may
be expressed by the angles $(\theta,\phi)$ as 
\begin{eqnarray}
\langle \vec S_{i,j,k} \rangle =  m\left(
\begin{array}{ccc}
(-1)^{i}\cos \theta \sin \phi \\  
(-1)^{i+j+k}\cos \theta \cos \phi \\  
\sin \theta 
\end{array}
\right). 
\end{eqnarray}
Thus in order to apply the HPT to the present case,
we use the representation
\begin{eqnarray}
\label{eq3-2}
\vec S_{i,j,k} &\rightarrow&  R_{x}\Big((-1)^{i+j+k}\theta\Big)\times  \nonumber \\
&&R_{z}\Big((-1)^{j+k}(-\phi)+\delta_{i+j+k,\rm{odd}}\,\,\pi\Big)
\vec S_{\rm HP},
\end{eqnarray}
where the operator $R_{\alpha}(\beta)$ represents
a rotation about $\alpha$ axis by
angle $\beta$. 
\begin{figure}
\scalebox{0.5}{\includegraphics{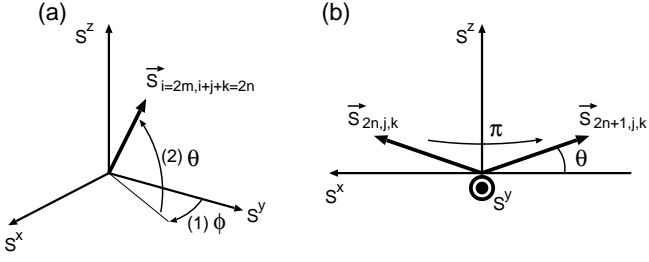}}
\caption{Definitions of the canting parameter.
In the SSB phase, as in (a), each spin direction is given by
two rotations (1) $R_{z}(-\phi)$ and (2) $R_{x}(\theta)$
applied in this order to spin pointing to $y$ direction.
In the symmetric phase or in the non-competitive case,
spins are canted in the $zx$ plane as shown in (b).
}
\label{Fig7}
\end{figure}
The operation (\ref{eq3-2}) is described in Fig.~\ref{Fig7}. 
For the original spin model (\ref{eq1-1}), these rotations 
correspond to a unitary transformation.
After these transformations, the leading order terms in $1/S$ is retained
to give a solvable Hamiltonian which is quadratic in bosonic operators.

The canting angles of the classical ground state
are given by minimizing the classical Hamiltonian, which is equivalent
to the MFT energy, Eq.~(\ref{eq:MFTenergy}). 
The canting angles in the classical ground state are thus given as
\begin{eqnarray}
\label{eq3-7}
\begin{array}{ll}
\cos\phi_{\rm cl} \sin\theta_{\rm cl} = \frac{H}{4S(J+J_{\perp}+J_{\perp}')},\\ 
\sin\phi_{\rm cl} = {\textstyle\frac{h}{4S(J_{\perp}+J_{\perp}')}}.
\end{array}
\end{eqnarray} 
For the resulting quadratic Hamiltonian 
$\hat H_{\rm HP}$, we perform a Fourier transformation (FT)
$c_{i,j,k}^{\dag}= N^{-d/2}\sum_{\vec k} 
{\rm e}^{i\vec k \cdot \vec r}c_{\vec k}^{\dag}$,
where $N$ is the linear system size, $d$ is the dimension of the system,
$k_{\alpha}$ $(|k_{\alpha}|<\pi/a_{\alpha})$ and $a_{\alpha}$ are,
respectively, the wave-number and the lattice constant for the
$\alpha$-direction.
A four-mode Bogoliubov
transformation (BT),~\cite{Colpa-BT}
which mixes
$c_{\vec k}, c_{-\vec k}, c_{\vec k-\vec\pi}, c_{-\vec k-\vec\pi}$
and their Hermitian conjugates ($\vec\pi\equiv(0,\pi/a_{y},\pi/a_{z})$), 
leads to the diagonalized form
\begin{eqnarray}
\label{eq3-9}
\hat H_{\rm HP} &=& \sum_{\vec k}\omega(\vec k)\,
\tilde{c}_{\vec k}^{\dag}\tilde{c}_{\vec k}+{\rm const},
\end{eqnarray}
with a single band in the first Brillouin zone,
where $\tilde{c}_{\vec k}$ is the magnon annihilation operator and 
the $k_{x}$ direction corresponds to the strongly AF one. 
The explicit results on the dispersion $\omega(\vec{k})$ are lengthy
and thus are given in Appendix~\ref{App2}.
Here we discuss physical implications of our results.

\begin{figure}
\scalebox{1.0}{\includegraphics{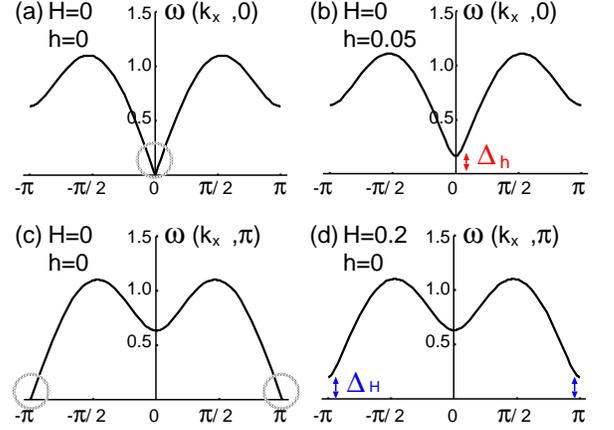}}
\caption{Magnon dispersions of the SSB phase in 
the 2D competitive case for $(S,J,J_{\perp})=(1/2,1,0.1)$ and $a_{x}=a_{y}=1$. 
The gray circles represent the gapless points.}
\label{Fig8}
\end{figure}

First, the obtained dispersion satisfies $\omega(\vec k)\geq 0$ for
all values of parameters.
This implies that
the SSB phase, which appears as the classical ground state, 
is stable against quantum fluctuations, at least in the lowest order of $1/S$.
Some representatives of the dispersion are shown in Fig.~\ref{Fig8}.
At zero field ($H=h=0$), there are two gapless points 
$\vec k_{h}=(0,0,0)$ and 
$\vec k_{H}=(\pi/a_{x},\pi/a_{y},\pi/a_{z})$, with linear dispersions
in the neighborhoods.
Let us define $\Delta_{h}=\omega(\vec k_{h})$ and
$\Delta_{H}=\omega(\vec k_{H})$.
Since our results in Appendix~\ref{App2-1} indicates
that $\Delta_{h}$ ($\Delta_{H}$) is non-vanishing
only when $h\neq 0$ ($H\neq 0$), we call it as
$h(H)$-induced gap. The true excitation gap, namely, the minimum
excitation energy, is given by $\Delta={\rm min}(\Delta_{h},\Delta_{H})$. 
Thus the gap $\Delta$ vanishes exactly 
as long as either $h$ or $H$ remains zero. This is contrasting to the
$S=1/2$ 1D HAF model, where the staggered field alone induces the 
gap.~\cite{Note2} The gapless excitations are identified as Nambu-Goldstone 
(NG) modes. Indeed, when either $H$ or $h$ is zero, the Hamiltonian 
has a continuous U(1) symmetry, which is broken spontaneously in the SSB
phase.

In the non-competitive case, 
the magnon dispersion relation
$\tilde\omega(\vec k)$ is given in Eq.~(\ref{eqB-2-8}) in Appendix~\ref{App2-2}.
Now the staggered field alone can open the gap, because the ground state
does not break any continuous symmetry spontaneously.
On the other hand, the system remains gapless
at $\vec{k}=0$ due to the NG mechanism if $h=0$ and $H$ is not too large.
The $h$-induced gap is thus defined as 
$\tilde \Delta_{h}=\tilde\omega(\vec k=\vec 0)$.

\begin{figure}
\scalebox{0.8}{\includegraphics{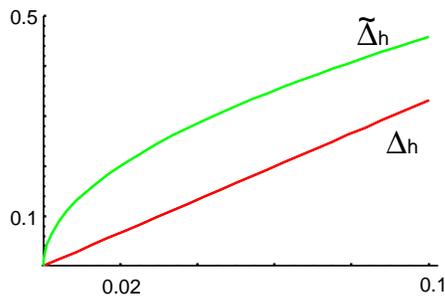}}
\caption{$\Delta_{h}$ and $\tilde{\Delta}_{h}$ in the 2D case for 
$(S,J,J_{\perp},H)=(1/2,1,0.1,0)$. 
In the present case where $H=0$, the gaps have simple forms: 
$\Delta_{h}=(1+J/J_{\perp})^{1/2} h$ and 
$\tilde{\Delta}_{h}=[1+h/(4SJ)]^{1/2}\sqrt{4SJh}$ (see Appendix).}
\label{Fig9}
\end{figure}

While the $h$-induced gap is defined for both cases,
there is a characteristic difference in the $h$ dependence of
the gaps. In the limit of small $h$, 
$\tilde \Delta_{h}\sim h^{1/2}$ for the non-competitive case,
but $\Delta_{h}\sim h$ for the competitive case.
In other words, one can say that the $h$-induced gap in the competitive case 
opens more slowly than in the non-competitive case. 
This is naturally understood because the competition between 
$J_{\perp}$ (or $J_{\perp}'$) and $h$ weakens the effect of the external 
symmetry breaking by $h$. It also implies that the ground state is more
stable in the non-competitive case, against quantum and thermal fluctuations.  
The opening gaps are drawn in Fig.~\ref{Fig9}. 
On the other hand, in the limit of small $H$, $\Delta_{H}\sim H$. 
Similarly to the case of $\Delta_{h}$,
this may be interpreted as a result of the competition 
between the uniform field and the AF couplings.

We expect the spin-wave theory for the gaps to be
qualitatively correct even for $S=1/2$.
As discussed in Appendix~\ref{App2-2}, 
the spin-wave dispersion of the symmetric phase can be obtained
by the replacement
$(|J_{\perp}|,|J_{\perp}'|)\rightarrow(-J_{\perp},-J_{\perp}')$ in the 
dispersion of the non-competitive case.

Finally let us discuss the relation of the present results to the 
spin-wave theory in the 1D model ($J_{\perp}=J_{\perp}'=0$) 
with the staggered field $h$ discussed in Ref.~\onlinecite{A-O}. The
spin-wave dispersion $\tilde\omega(\vec k)$ for the non-competitive case 
does approach smoothly to the 1D dispersion [Eq.~(3.11) in
Ref.~\onlinecite{A-O}], when $J_{\perp},J_{\perp}'\rightarrow 0$.
On the other hand, the dispersion $\omega(\vec k)$ for the competitive
case apparently does not reduce to the 1D one in the limit of 
$J_{\perp},J_{\perp}'\rightarrow 0$. This is because $\omega(\vec k)$ is 
the dispersion of the magnon excitation in the SSB phase, which is
absent in the 1D model. In fact, for any finite $h$, the SSB phase 
is realized [and hence $\omega(\vec k)$ is applicable] only when 
$J_{\perp}$ and $J_{\perp}'$ are above the critical values.
Thus by decreasing  $J_{\perp}(J_{\perp}')$ at a fixed $h$, the 
system undergoes a phase transition into the symmetric phase.
At the transition the dispersion should also change drastically.
In the symmetric phase, similarly to the non-competitive case, 
the dispersion does approach continuously to the 1D dispersion.

\section{Chain Mean-Field Theory Approach}
\label{Sec4}
In this section, we reconstruct the phase diagram of our model 
using chain mean-field theory (CMFT). In the CMFT,
weak couplings among the chains are treated with a MFT, and
the resulting effective 1D problem is analyzed as precisely as possible. 
If the 1D problem can be treated exactly, the CMFT is expected to be much 
reliable when the one dimensionality is strong enough as
in the case of Cu-benzoate, 
since it includes the fluctuations in the strongly coupled direction correctly.
The usefulness of the CMFT has been demonstrated in several
applications.~\cite{CMF-Scal,CMF-Sch,CMF-Wang} 

In Sec.~\ref{Sec4-1}, we discuss how the CMFT determines the phase transition
for our model (\ref{eq1-1}). Sec.~\ref{Sec4-2} is a brief overview 
of susceptibilities of the $S=1/2$ HAF chain which
are necessary to the CMFT. In Sec.~\ref{Sec4-3}, we present the CMFT
phase diagrams and compare them to the MFT ones.

\subsection{CMFT for Our Model}
\label{Sec4-1}
Let us derive the effective 1D model for our system (\ref{eq1-1}),
within the CMFT.

In the competitive case, we consider the symmetric phase side 
for convenience. The mean-field procedure for the weak inter-chain couplings
replaces them with the effective external fields.
Thus the resulting Hamiltonian is
\begin{eqnarray}
\label{eq4-1-1}
\hat H &\rightarrow& \sum_{j,k}\hat H_{j,k}
+N^{d}(J_{\perp}+J_{\perp}')(m_{y}^{2}-m_{x}^{2}-m_{z}^{2}),\nonumber\\
\hat H_{j,k}&=& \sum_{i}J \vec S_{i,j,k} \cdot \vec S_{i+1,j,k}\nonumber\\
&&-(-1)^{j+k}[h'+2(J_{\perp}+J_{\perp}')m_{y}](-1)^{i}S_{i,j,k}^{y}\nonumber\\
&&-[h-2(J_{\perp}+J_{\perp}')m_{x}](-1)^{i}S_{i,j,k}^{x}\nonumber\\
&&-[H-2(J_{\perp}+J_{\perp}')m_{z}]S_{i,j,k}^{z},
\end{eqnarray}
where we introduced an infinitesimal staggered field $(-1)^{i+j+k}h'$ parallel
to the order parameter. 
The CMFT requires that the mean fields $m_{x,y,z}$ are equivalent 
to the corresponding moments of the effective chain $\hat H_{j,k}$. 
It has the two effective staggered fields 
$h_{x}\equiv h-2(J_{\perp}+J_{\perp}')m_{x}$ and 
$h_{y}\equiv h'+2(J_{\perp}+J_{\perp}')m_{y}$ as well as the
effective uniform field $H_{z}\equiv H-2(J_{\perp}+J_{\perp}')m_{z}$.
Clearly it is sufficient to consider one chain where $j+k$ is even, and 
we represent it as $\hat H_{\rm 1D}$. 
Within the linear-response theory, the moment $|\langle S_{i}^{y} 
\rangle |$, in which $\langle \cdots\rangle$ stands for the mean value of
$\hat H_{\rm 1D}$, can be approximated by 
$\chi_{y}^{\rm 1D}(H_{z},h_{x}, 0)h_{y}$ where the staggered susceptibility
is defined as $\chi_{y}^{\rm 1D}(H_{z},h_{x}, h_{y})\equiv 
\frac{\partial |\langle S_{i}^{y}\rangle |}{\partial h_{y}}$.
Therefore the above requirement leads to $m_{y}=\chi_{y}^{\rm 1D}h_{y}$ 
which is transformed to
\begin{eqnarray}
\label{eq4-1-2}
m_{y} &=& {\textstyle \frac{\chi_{y}^{\rm 1D}}
{1-2(J_{\perp}+J_{\perp}')\chi_{y}^{\rm 1D}}\,\,h'}.
\end{eqnarray}
Similarly $m_{x}$ and $m_{z}$ can be determined by the CMFT as well. 
The critical condition of our model is 
\begin{eqnarray}
\label{eq4-1-3}
1-2(J_{\perp}+J_{\perp}')\,\,\chi_{y}^{\rm 1D}(H_{z},h_{x},0)= 0.
\end{eqnarray}
At this point, one sees that the original 3D or 2D susceptibilities and
moments can be described by those of the effective chain.

The staggered field in $\hat H_{\rm 1D}$ has both $x$ and
$y$ components: $h_{x}$ and $h_{y}$. By the infinitesimal rotation 
$R_{z}(\delta)$ about $z$ axis by angle $\delta$, where 
$\sin\delta =-[h_{y}^{2}/(h_{y}^{2}+h_{x}^{2})]^{1/2}$, the 
effective Hamiltonian is simplified as 
\begin{eqnarray}
\label{eq4-1-6}
\hat H_{\rm 1D}'= \sum_{i}J\vec S_{i}'\cdot \vec S_{i+1}'
-H_{z}{S_{i}^{\prime}}^{z} -h_{x}'(-1)^{i}{S_{i}^{\prime}}^{x},
\end{eqnarray}
with the one-component staggered field $h_{x}'= h_{x}/\cos\delta$. 
In order to obtain an alternative formula determining $m_{y}$ instead of
Eq.~(\ref{eq4-1-2}), we focus on the relation  
\begin{eqnarray}
\label{eq4-1-7}
\langle {S_{i}^{\prime}}^{x} \rangle' &=& \cos\delta \,\,
\langle S_{i}^{x} \rangle 
-\sin\delta \,\,\langle S_{i}^{y} \rangle \nonumber\\
&=& (-1)^{i}(\cos\delta \,\,m_{x}- \sin\delta \,\,m_{y}),
\end{eqnarray}
where $\langle \cdots\, \rangle'$ represents the expectation value of 
$\hat H_{\rm 1D}'$ and the second equality is caused by the 
self-consistency of the CMFT. 
Here we define the susceptibilities of the HAF chain (\ref{eq4-1-6}) as 
$\chi_{u}^{\rm 1D}(T,H_{z},h_{x}')\equiv\frac{\partial m_{z}'}{\partial H_{z}}$ and 
$\chi_{s}^{\rm 1D}(T,H_{z},h_{x}')\equiv\frac{\partial m_{x}'}{\partial
h_{x}'}$ where $m_{x,z}'=|\langle {S_{i}^{\prime}}^{x,z}\rangle'|$.
Within the linear-response theory Eq.~(\ref{eq4-1-7}) is reduced to
\begin{eqnarray}
\label{eq4-1-8}
\chi_{s}^{\rm 1D}(T,H_{z},h_{x}')\,\,h_{x}'
=\cos\delta \,\,m_{x}-\sin\delta \,\,m_{y}.
\end{eqnarray}
Let us recall that $\delta$ is defined by $m_{x}$ and $m_{y}$, and 
that $m_{x,z}$ are determined by the CMFT. 
Consequently, $m_{y}$ can be determined as a solution to Eq.~(\ref{eq4-1-8}). 
The condition that $m_y$ diverges would determine the phase transition.

\subsection{Susceptibilities of $S=1/2$ AF chains}
\label{Sec4-2}
In order to solve Eqs.~(\ref{eq4-1-3}) or (\ref{eq4-1-8})
in terms of $m_y$, we need 
the explicit forms of the susceptibilities: $\chi^{\rm 1D}_{u}$ and
$\chi^{\rm 1D}_{s}$.
Here we briefly summarize the known results~\cite{A-O,Aff,Boso, Gogo}
on these quantities, obtained by the bosonization technique.

In the absence of the staggered field $h'_x$, the low-energy effective
theory of of the Heisenberg chain~(\ref{eq4-1-6}) is given by a
free boson field theory, and the uniform susceptibility
at zero temperature is obtained as
\begin{eqnarray}
\label{eq4-2-6}
\chi_{u}^{\rm 1D}(T,H,0)&\approx&\frac{a}{(2\pi)^{2}R(H)^{2}v(H)},
\end{eqnarray}
where $H_z=H$, and
$R$ and $v$, respectively, are the compactification radius of the effective boson
field theory~\cite{Aff,Boso,Gogo} and the spin-wave velocity.
The exact values $v(H)$ and $R(H)$ as functions of
the uniform field $H$ are given by a solution of
a set of the Bethe ansatz integral equations.~\cite{K-B-I} 
In the case of $H=0$, they are explicitly given as
\begin{eqnarray}
\label{eq4-2-5}
v= \pi \,J\,a/2,&&
R=1/\sqrt{2\pi} .
\end{eqnarray}
For a small uniform field $H_z=H (\ll J)$, 
the asymptotic behavior of the radius $R$ follows~\cite{A-O}
\begin{eqnarray}
\label{eq4-2-5*}
2\pi R^{2}&\approx& 1-\frac{1}{2\ln(J/H)}. 
\end{eqnarray}
The logarithmic temperature correction due to the marginal operator
of the HAF chain is discussed in Ref.~\onlinecite{E-A-T}.
Although the transverse staggered field $h'_x$ induces a gap, it
is expected to have little effect~\cite{O-A,A-O} on the uniform
susceptibility $\chi^{\rm 1D}_u$ if $h'_x$ is small enough.

Next we turn to the staggered susceptibility $\chi_{s}^{\rm 1D}$. 
In the case $J \gg h'_x,\, k_{B}T$, the low-energy effective theory
of the Hamiltonian~(\ref{eq4-1-6}) is given by a quantum
sine-Gordon field theory.
Using the exact solutions of the HAF chain, the scaling arguments and 
Lukyanov-Zamolodchikov prediction~\cite{Luk-Zam,Lukya}, 
the staggered susceptibility $\chi_{s}^{\rm 1D}$ 
for small $H_z=H$ and $h'_x=h$ at $T=0$~\cite{O-A,A-O} is given as
\begin{widetext}
\begin{eqnarray}
\label{eq4-2-9-1}
\chi_{s}^{\rm 1D}&\approx&\left\{
\begin{array}{lll}
D[2(1-2\pi R^{2})]^{-1/3}
\frac{\pi R^{2}}{2-\pi R^{2}}
\left(\frac{J}{H}\right)^{-2(1-2\pi R^{2})/3}
\left(\frac{h}{J}\right)^{\pi R^{2}/(2-\pi R^{2})}\,h^{-1}& (H \gg h)\\
\label{eq4-2-9-2} 
D\,e^{-1/3}\frac{1}{3}\left[\ln\left(\frac{J}{H}\right)\right]^{1/3}
\left(\frac{h}{J}\right)^{1/3}\,h^{-1}&(J\gg H\gg h)\\
\label{eq4-2-9-3}
D\frac{2^{1/3}}{3^{4/3}}
\left[\frac{h}{J}\ln\left(\frac{J}{h}\right)\right]^{1/3}\,h^{-1}&(H=0)
\end{array}
\right.,
\end{eqnarray}
\end{widetext}
where the radius $R$ is that of the model without the staggered field,
and $D\equiv 0.3868\dots $.
The first formula is actually valid for $H \leq 2J$ (below saturation field),
but the second is only so for $H \ll J$.
These formulas are correct in $k_{B}T\ll h \ll J$. (More precisely, 
within $k_{B}T\ll \Delta_{h} \ll J$ where $\Delta_{h}$ is the 
$h$-induced gap in the HAF chain.~\cite{A-O}) 

In the intermediate temperature regime $h,H\ll k_{B}T\ll J$, 
where the temperature is larger than the induced gap,
the staggered susceptibility may be approximated by that for zero
staggered field~\cite{A-O} as
\begin{eqnarray}
\label{eq4-2-10}
 \chi_{s}^{\rm 1D}(k_{B}T\gg h)&\approx& {\cal D}\,\, 
\frac{[\ln(J/k_{B}T)]^{1/2}}{k_{B}T},
\end{eqnarray}   
where ${\cal D}\equiv 0.2779\dots$.

\subsection{Phase diagrams in CMFT}
\label{Sec4-3}
Employing the results of Secs.~\ref{Sec4-1} and \ref{Sec4-2}, 
we study the phase diagrams, in particular for the competitive case,
within the CMFT. Unfortunately, it can be applied only to several limited
regions in the parameter space, where the susceptibilities of the 1D
model are obtained exactly.

\begin{figure}
\scalebox{0.9}{\includegraphics{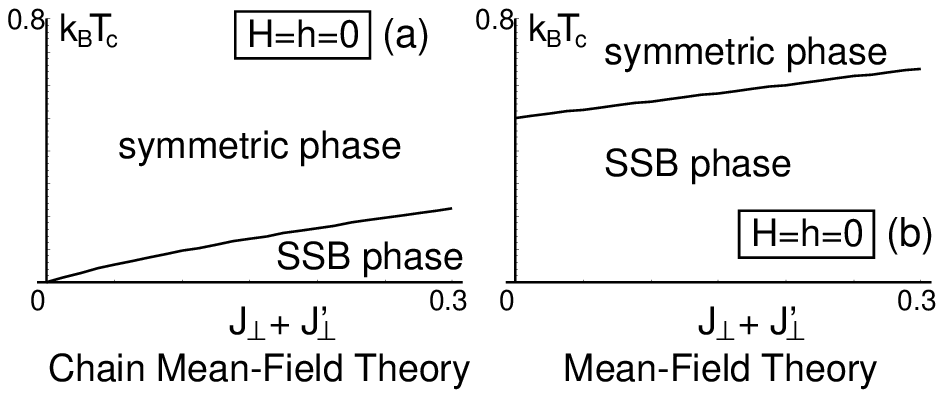}}
\caption{$(k_{B}T_{c},J_{\perp}+J_{\perp}')$ in the
 competitive case for $J=1$. (a) and (b) are, respectively, 
predicted within the CMFT and the MFT. It is confirmed that the SSB areas of 
the CMFT are considerably smaller than the MFT.}
\label{Fig10}
\end{figure}

First, let us consider the region near $(k_{B}T_{c},0,0)$. In the 
zero-field case, the effective model $\hat H_{\rm 1D}$ has only
an infinitesimal staggered mean field $h_{y}$ 
as long as it is in the symmetric phase. Therefore the formula
(\ref{eq4-2-10}) can give the self-consistent value of $m_{y}$, and 
Eq.~(\ref{eq4-1-3}) is the critical condition which serves 
the critical temperature:
\begin{eqnarray}
\label{eq4-3-2}
\begin{array}{lll}
k_{B}T_{c} &=& 2{\cal D}\,\,(J_{\perp}+J_{\perp}') 
\left[\ln \left(\frac{J}{k_{B}T_{c}}\right)\right]^{1/2}\\
&\approx & 2{\cal D}\,\,(J_{\perp}+J_{\perp}')
\left[\ln \left(\frac{J}{2{\cal D}\,\,(J_{\perp}+J_{\perp}')}
\right)\right]^{1/2},
\end{array}
\end{eqnarray}
which is reasonable when $k_{B}T_{c},\,(J_{\perp}+J_{\perp}')\ll J$ 
and agrees with the result of Ref.~\onlinecite{CMF-Sch}. 
Equation (\ref{eq4-3-2}) is also stable against a small staggered field $h$,
because $\chi^{\rm 1D}_{s}$ is independent of $h$ in Eq.~(\ref{eq4-2-10}).
These results are also valid for the non-competitive case, with the 
replacement 
$J_{\perp}+J_{\perp}'\rightarrow |J_{\perp}|+|J_{\perp}'|$. 
Figure \ref{Fig10} represents the comparison between the CMFT result
(\ref{eq4-3-2}) and the mean-field prediction: $k_{B}T_{c}=(J+J_{\perp}+J_{\perp}')/2$. 
\begin{figure}
\scalebox{0.9}{\includegraphics{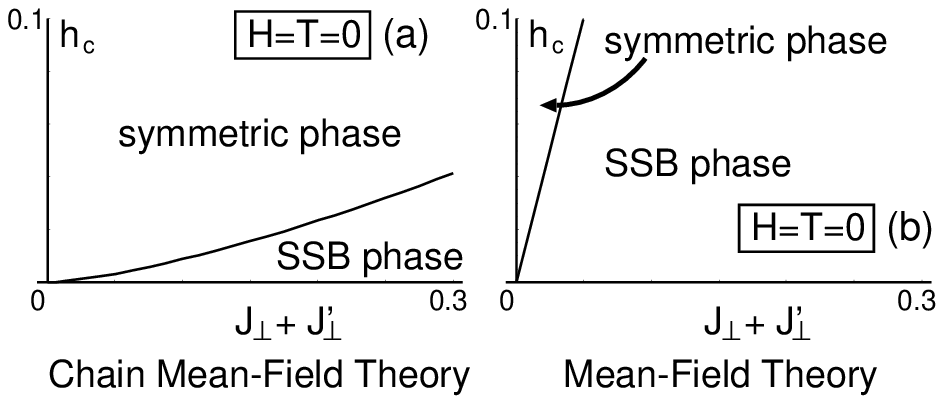}}
\caption{$(h_{c},J_{\perp}+J_{\perp}')$ in the
 competitive case for $J=1$. (a) and (b) are, respectively, 
predicted within the CMFT and the MFT. Similarly to Fig.~\ref{Fig10}, 
the narrowing of the SSB areas occurs.}
\label{Fig10-2}
\end{figure} 

Next we investigate the neighborhood of $(0,0,h_{c})$. 
In this case ($H=0$), $\hat H_{\rm 1D}$ has two staggered fields. 
Therefore through the rotation of Sec.~\ref{Sec4-2}, the order parameter 
$m_{y}$ can be fixed by Eq.~(\ref{eq4-1-8}). 
Inserting the third formula of Eq.~(\ref{eq4-2-9-3}) into Eq.~(\ref{eq4-1-8}), 
and performing the Taylor expansion of Eq.~(\ref{eq4-2-9-3}) around
$h_{y}=0$, we obtain the linear-response relation:
\begin{eqnarray}
\label{eq4-3-3}
\begin{array}{lll}
m_{y}=\frac{\left\{\frac{2}{3}-\frac{1}{6}[\ln(J/h_{x})]^{-1}
\right\}m_{x}}{h-\frac{1}{3}\left\{10(J_{\perp}+J_{\perp}')
-[\ln(J/h_{x})]^{-1}\right\}m_{x}}\,\,h'.\\
\end{array}
\end{eqnarray}
Therefore the critical condition can be written as
\begin{eqnarray}
\label{eq4-3-4}
m_{x} &\approx& \frac{3\,h_{c}}{10(J_{\perp}+J_{\perp}')}, 
\end{eqnarray}
where we assumed $h_{c},\,(J_{\perp}+J_{\perp}')\ll J$. 
On the other hand, $m_{x}$ should be fixed from 
the self-consistency of the CMFT as well.
At $T=0$, it leads to
\begin{eqnarray}
\label{eq4-3-5} 
\begin{array}{lll}
m_{x} = \frac{\chi_{s}^{\rm 1D}\left(0,0,h-2(J_{\perp}+J_{\perp}')m_{x}\right)}
{1+2(J_{\perp}+J_{\perp}')
\chi_{s}^{\rm 1D}\left(0,0,h-2(J_{\perp}+J_{\perp}')m_{x}\right)}\,\,h. 
\end{array}
\end{eqnarray}
Combining Eqs.~(\ref{eq4-3-4}) and (\ref{eq4-3-5}),
we obtain the critical staggered field:
\begin{eqnarray}
\label{eq4-3-6}
\begin{array}{lll}
h_{c}&=& D'\left[\frac{J_{\perp}+J_{\perp}'}{J}
\ln\left(\frac{5J}{2h_{c}}\right)\right]^{1/2}2(J_{\perp}+J_{\perp}') \\
&\approx & D' \left[\frac{J_{\perp}+J_{\perp}'}{J}
\ln \left(\frac{5}{4D'}\left(\frac{J}{J_{\perp}+J_{\perp}'}\right)
^{1/2}\right)\right]^{1/2}\\
&&\times 2(J_{\perp}+J_{\perp}'), 
\end{array}
\end{eqnarray} 
where we kept only the leading order of $(J_{\perp}+J_{\perp}')/J$ and 
$D'\equiv (2\times 10^{2}D^{3}3^{-7})^{1/2}\approx 7.27\dots
\times 10^{-2}$. This is valid if $h_{c},\,(J_{\perp}+J_{\perp}')\ll J$,
and is compared to the MFT result $h_{c}=2(J_{\perp}+J_{\perp}')$
extracted from Eq.~(\ref{eq2-5}) in Fig.~\ref{Fig10-2}. 
The CMFT correction to the MFT is found as 
a significant multiplicative factor $D'[\cdots]^{1/2}$.

\begin{figure}
\scalebox{0.8}{\includegraphics{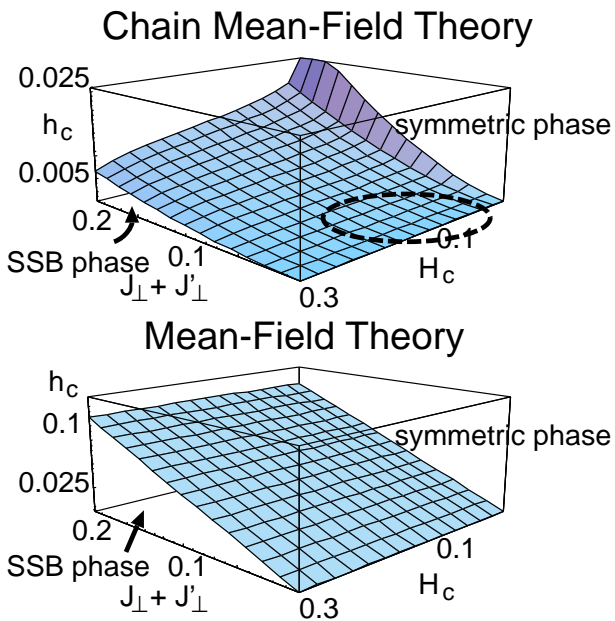}}
\caption{Critical surface $(h_{c},H_{c})$ in the competitive case for
 $(J,T)=(1,0)$ obtained with the CMFT (\ref{eq4-3-8}) (upper panel) and 
the MFT (\ref{eq2-5}) (lower panel). 
In the whole region of the CMFT panel, the critical region surrounded by 
the dashed ring ($h_{c}\ll H_{c}\ll J$) is greatly reliable.} 
\label{Fig11}
\end{figure}

We consider furthermore the critical line in the $T=0$ plane
in the limit of $h_{c}\ll H_{c}\ll J$. In the symmetric phase 
near this line, the effective model $\hat H_{\rm 1D}$ has all three
kinds of the external fields. Hence $m_{x}$ and $m_{z}$ must be determined
concurrently by the CMFT scheme. However in the present case $h_{c}\ll H_{c}$,
$m_{x}$ may be estimated independently by taking an approximation 
$\chi_{u}^{\rm 1D}(0,H,h)\sim \chi_{u}^{\rm 1D}(0,H,0)$ 
which was given in Eq.~(\ref{eq4-2-6}). From this approximation,
$m_{z}$ is also fixed as 
\begin{eqnarray}
\label{eq4-3-7}
m_{z}&\approx&\frac{H}{{\textstyle\pi^{2}J(1-[\ln(J/H)]^{-1})
+2(J_{\perp}+J_{\perp}')}},
\end{eqnarray}
which is justified in $H,\,(J_{\perp}+J_{\perp}')\ll J$. 
In the sufficiently small field case, $J\gg H(\gg h)$, 
the logarithmic part can be dropped as well. 
From the second of Eq.~(\ref{eq4-2-9-1}) and Eqs.~(\ref{eq4-3-4}), 
(\ref{eq4-3-5}) and (\ref{eq4-3-7}), we obtain the critical line
in $T=0$
\begin{eqnarray}
\label{eq4-3-8}
h_{c}&=& D''\Big[
\begin{array}{ll}
\frac{J_{\perp}+J_{\perp}'}{J}
\Big(\ln\Big(\frac{\pi^{2}J (1-[\ln(J/H_{c})]^{-1})
+2(J_{\perp}+J_{\perp}')}{\pi^{2}H_{c}}\Big)
\end{array}\nonumber\\
&&+[\ln(J/H_{c})]^{-1}+\cdots \,\Big)\Big]^{1/2}\,2(J_{\perp}+J_{\perp}'),
\end{eqnarray} 
where $D''\equiv (10^{2}D^{3}e^{-1}3^{-6})^{1/2}\approx 5.40\dots\times 10^{-2}$.
This condition is presumably suitable for $k_{B}T_{c}\ll h_{c}\ll H_{c}\ll J$.
Thus, in order to obtain a more precise condition,
it is necessary to adopt the first
formula (\ref{eq4-2-9-1}) and the exact values of $R$ and $v$.
Figure \ref{Fig11} exhibits the comparison between this line and the
mean-field prediction.

Finally, we consider the point $(0,H_{c},0)$, where 
$\hat H_{\rm 1D}$ has only the uniform field 
$H_{z}=H_{c}-2(J_{\perp}+J_{\perp}')m_{z}$ except for the
infinitesimal field $h_{y}$. It has been known from Bethe ansatz that
the magnetization saturates at the point $H_{z}=2J$ at $T=0$ in the HAF
chain having only the field $H_{z}$. 
The transition point between the SSB and symmetric phase 
in this case should be identified with the
saturation of the uniform magnetization. 
Hence, within the CMFT, the critical uniform field is given as 
\begin{eqnarray}
\label{eq4-3-9}
H_{c}&=& 2J + (J_{\perp}+J_{\perp}'). 
\end{eqnarray} 
The substitution $J_{\perp}+J_{\perp}'\rightarrow
-(|J_{\perp}|+|J_{\perp}'|)$ gives the critical field of the 
AF-paramagnetic transition in the non-competitive case.

\begin{figure}
\scalebox{0.6}{\includegraphics{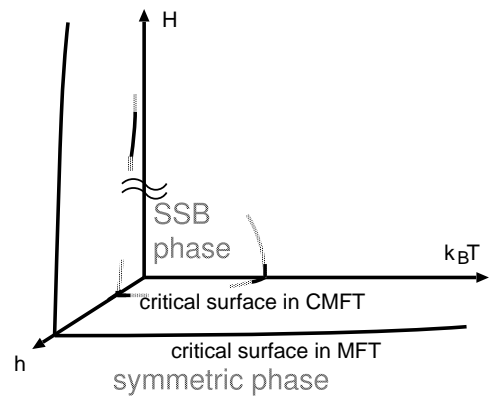}}
\caption{Schematic phase diagrams of the CMFT and the MFT 
in the competitive case.}
\label{Fig12}
\end{figure}

From these results, we can compare the CMFT and the MFT phase diagrams.
The comparison for the competitive case is summarized in Fig.~\ref{Fig12}.
In the case of weak inter-chain couplings ($J\gg J_{\perp},J_{\perp}'$),
the SSB phase of
the CMFT is much smaller than one of the MFT. Especially there is a
significant narrowing in $k_{B}T$ and $h$ directions. 
As seen from Eq.~(\ref{eq4-1-2}), this is because the phase transition in the
CMFT framework is driven by the divergence of the susceptibility
(in the present case, $\chi_{s}^{\rm 1D}$) in the effective chain, 
while temperature and the field $h_{x}$ (or $m_{x}$) suppress the divergence.
On the other hand, the reduction of the critical uniform field $H_{c}$ is
small in the weakly coupled case. This is because the uniform field 
competes with the intra-chain AF interaction as well as with 
the inter-chain interactions.
  
Finally we review the validity of the two theories. Since the strong 
one dimensionality is the basis of the CMFT procedure, it is
expected that the CMFT is more reliable in the limit of
$J\gg J_{\perp},J_{\perp}'$. 
On the other hand, when $J_{\perp}$ and $J_{\perp}'$ are comparable to
$J$, the special treatment of only one direction is unjustified. 
Therefore the MFT, which treats all couplings equally, is more reasonable for 
$J\sim J_{\perp},J_{\perp}'$.  

\section{Summary and Discussion}
\label{Sec5}
We considered the effects of the staggered field $h$ in an $S=1/2$ Heisenberg
antiferromagnet (\ref{eq1-1}) in two or three dimensions.
The system behaves quite differently depending on whether the staggered
field and the inter-chain couplings are competitive or not.
In the competitive case, the appearance
of a characteristic ordered (SSB) phase is predicted by the
MFT. The SSB phase breaks the translational symmetry of the weakly
coupled direction, and therefore it is peculiar to high dimensional systems. 
We also applied the CMFT to the model (\ref{eq1-1}), and predicted that 
the region of the SSB phase becomes narrow in the CMFT scheme.  
The MFT and the CMFT are valid, respectively, in 
$J \sim J_{\perp},J_{\perp}'$ and in $J \gg J_{\perp},J_{\perp}'$. 
The crossover behavior between these two regions can not
be described by the mean-field type approach.~\cite{A-G-S,Aff-Halp,MCMF-Sand}
It would require a more precise treatment of fluctuations.

Moreover we studied spin-wave theory in both the competitive 
and non-competitive cases at $T=0$.
When the uniform field $H$ is non-vanishing, the $h$-induced gap opens
as $\Delta_{h} \sim h$ 
in the SSB phase, while $\tilde\Delta_{h} \sim h^{1/2}$ in the 
AF phase of the non-competitive case. This difference reflects the partial
cancellation of the staggered field effect due to the competitive
inter-chain interaction in the SSB phase. 
The spin-wave dispersion in the SSB phase remains gapless due to NG
mechanism even under a non-vanishing $h$. 
This is in contrast to the case of the 1D model.

Finally we comment on a few recent reports related to our study.
In $\rm BaCu_{2}Si_{2}O_{7}$ reported in Ref.~\onlinecite{BaCuSi},
both the staggered field and 
the inter-chain interactions are expected, as in our models. 
However, the effect of the exchange anisotropies, which is ignored in
the present paper, is argued to be responsible for the observed two
spin-flop transitions. Extending the present work to such a system would
be an interesting problem in the future. Furthermore,
in $\rm BaCu_{2}(Si_{1-x}Ge_{x})_{2}O_{7}$,~\cite{BaCu2-x} the sign of the
inter-chain interaction seems to depend on the doping parameter x.
Thus, it could provide a realization of the competitive and
non-competitive cases.

Wang {\em et al}.~\cite{Ladd-Wang} investigated an $S=1/2$ AF ladder 
system with a staggered field. 
They argue that the competition between the staggered field and the
rung interaction brings a quantum criticality. It might be interesting
to compare our analysis on the higher dimensional system with theirs.

\section*{Acknowledgements}
We thank Collin Broholm, Dan Reich and Hidekazu Tanaka for useful
comments.
This work was partially supported by Grant-in-Aid by MEXT of Japan.

\appendix
\section{Details of Spin-wave Results in Sec.~\ref{Sec3}}
\label{App2}
Here we supplement the spin-wave results omitted in Sec.~\ref{Sec3}.
\subsection{The competitive case}
\label{App2-1}
We write down the details of the competitive case. 
After the FT of the boson operator $c_{i,j,k}$, 
the spin-wave Hamiltonian can be expressed as the following matrix form:
\begin{eqnarray}
\label{eqB-1-1}
\hat H_{\rm HP}&=&
\sum_{\vec k}\left(
\begin{array}{ll}
{\cal C}_{\vec k}^{\dag\,\,T}&{\cal C}_{\vec k}^{\,\,T}
\end{array}
\right)
\left(
\begin{array}{cc}
\xi_{\vec k} & \eta_{\vec k}\\
\eta_{\vec k}^{*}&\xi_{\vec k}^{*}
\end{array}
\right)
\left(
\begin{array}{l}
{\cal C}_{\vec k}\\
{\cal C}_{\vec k}^{\dagger}
\end{array}
\right)-4E_{1}(\vec k)\nonumber\\
&&+ E_{\rm gs}^{\rm cl},
\end{eqnarray}
where $*$ and $T$, respectively, stand for the complex conjugate of
each matrix component and the transpose of matrices, 
and the $4\times 4$ matrices $\xi_{\vec k}$, $\eta_{\vec k}$ 
and the column four-vector ${\cal C}_{\vec k}$ are given as
\begin{eqnarray}
\xi_{\vec k}&=&\left(
\begin{array}{cccc}
E_{1}(\vec k)&0&0& i E_{4}(k_{y,z})\\
0&E_{1}(\vec k)&i E_{4}(k_{y,z})&0\\
0&-iE_{4}(k_{y,z})&F_{1}(\vec k)&0\\
-iE_{4}(k_{y,z})&0&0&F_{1}(\vec k)
\end{array}
\right),\nonumber\\
\eta_{\vec k}&=&\left(
\begin{array}{cccc}
0&E_{2}(\vec k)&0& iE_{3}(k_{x})\\
E_{2}(\vec k)&0&iE_{3}(k_{x})&0\\
0&iE_{3}(k_{x})&0&F_{2}(\vec k)\\
iE_{3}(k_{x})&0&F_{2}(\vec k)&0
\end{array}
\right),\nonumber\\
{\cal C}_{\vec k}&=&\left(
\begin{array}{cccc}
c_{\vec k}&c_{-\vec k}&c_{\vec k-\vec\pi}&c_{-\vec k-\vec\pi}
\end{array}
\right)^{T}.
\end{eqnarray}
Here $E_{1,2,3,4}$ are defined as
\begin{widetext}
\begin{eqnarray}
E_{1}(\vec k)&=&  \Big[2SJ(1-2\sin^{2}\theta_{\rm cl}\cos^{2}\phi_{\rm cl})
+2S(J_{\perp}+J_{\perp}')(2\cos^{2}\theta_{\rm cl}\cos^{2}\phi_{\rm cl}-1)
-2SJ\sin^{2}\theta_{\rm cl}\cos^{2}\phi_{\rm cl}\cos k_{x}a_{x}\nonumber\\
&&+2S(\cos^{2}\theta_{\rm cl}(1+\sin^{2}\phi_{\rm cl})-1)
(J_{\perp}\cos k_{y}a_{y}+J_{\perp}'\cos k_{z}a_{z})
+H\sin\theta_{\rm cl}\cos\phi_{\rm cl}+h\sin\phi_{\rm cl} \Big]/8,\nonumber\\
E_{2}(\vec k)&=&\Big[ SJ(\cos^{2}\theta_{\rm cl}
-\sin^{2}\theta_{\rm cl}\sin^{2}\phi_{\rm cl})\cos k_{x}a_{x}
+S\cos^{2}\theta_{\rm cl}\cos^{2}\phi_{\rm cl}
(J_{\perp}\cos k_{y}a_{y}+J_{\perp}'\cos k_{z}a_{z})\Big]/4,\nonumber\\
E_{3}(k_{x})&=& -SJ\sin 2\theta_{\rm cl}\sin\phi_{\rm cl}\cos k_{x}a_{x}\,\,/4,\nonumber\\
E_{4}(k_{y,z})&=& S\sin 2\theta_{\rm cl}\sin\phi_{\rm cl}
(J_{\perp}\cos k_{y}a_{y}+J_{\perp}'\cos k_{z}a_{z})/4,
\end{eqnarray}
\end{widetext}
and $F_{1,2}(\vec k)=E_{1,2}(\vec k-\vec\pi)$.
Thus $\xi_{\vec k}$ is Hermitian and $\eta_{\vec k}$ is 
symmetric. Let us suppose that a four-mode BT 
\begin{equation}
\label{eqB-1-2}
\left(
\begin{array}{cc}
\tilde{\cal C}_{\vec k} \\
\tilde{\cal C}_{\vec k}^{\dag}
\end{array}
\right)= M_{\rm BT}(\vec k)
\left(
\begin{array}{cc}
{\cal C}_{\vec k} \\
{\cal C}_{\vec k}^{\dag}
\end{array}
\right),
\end{equation}
where $\tilde{\cal C}_{\vec k}=(\tilde{c}_{\vec k},\tilde{c}_{-\vec k},
\tilde{c}_{\vec k-\vec\pi},\tilde{c}_{-\vec k-\vec\pi})^{T}$ is a set of new
boson (magnon) operators and $M_{\rm BT}(\vec k)$ is an $8\times 8$ matrix,
diagonalizes the Hamiltonian as follows:
\begin{eqnarray}
\label{eqB-1-3}
\hat H_{\rm HP}&=&
\sum_{\vec k}\left(
\begin{array}{ll}
\tilde{\cal C}_{\vec k}^{\dag\,\,T}&\tilde{\cal C}_{\vec k}^{\,\,T}
\end{array}
\right)
\left(
\begin{array}{cc}
\Omega_{\vec k}& 0\\
0&\Omega_{\vec k}
\end{array}
\right)
\left(
\begin{array}{l}
\tilde{\cal C}_{\vec k}\\
\tilde{\cal C}_{\vec k}^{\dagger}
\end{array}
\right)-4E_{1}(\vec k)\nonumber\\
&&+ E_{\rm gs}^{\rm cl},
\end{eqnarray}
where $\Omega_{\vec k}={\rm diag}(\omega_{1}(\vec k),\omega_{2}(\vec k),
\omega_{3}(\vec k),\omega_{4}(\vec k))$. According to
Ref.~\onlinecite{Colpa-BT},
determining $\Omega_{\vec k}$ and $M_{\rm BT}(\vec k)$ is equivalent to
solving an eigenvalue problem
\begin{eqnarray}
\label{eqB-1-4}
\left(
\begin{array}{cc}
\xi_{\vec k} & -\eta_{\vec k}\\
\eta_{\vec k}^{*}&-\xi_{\vec k}^{*}
\end{array}
\right)M_{\rm BT}^{\dag}&=&
M_{\rm BT}^{\dag}
\left(
\begin{array}{cc}
\Omega_{\vec k}& 0\\
0&-\Omega_{\vec k}
\end{array}
\right).
\end{eqnarray}
The eight eigenvalues $\pm \omega_j(\vec{k})$ ($j=1,2,3,4 $) are
given by $\pm \lambda_l$ ($l=1,2,3,4$), where 
\begin{eqnarray}
\label{eqB-1-5}
\lambda_{1}(\vec k)&=&[1/2\, G_{\vec k}
-1/2(G_{\vec k}^{2}-4G_{\vec k}^{+})^{1/2}]^{1/2},\nonumber\\
\lambda_{2}(\vec k)&=&[1/2\, G_{\vec k}
-1/2(G_{\vec k}^{2}-4G_{\vec k}^{-})^{1/2}]^{1/2},\nonumber\\ 
\lambda_{3}(\vec k)&=&[1/2\, G_{\vec k}
+1/2(G_{\vec k}^{2}-4G_{\vec k}^{+})^{1/2}]^{1/2},\nonumber\\
\lambda_{4}(\vec k)&=&[1/2\, G_{\vec k}
+1/2(G_{\vec k}^{2}-4G_{\vec k}^{-})^{1/2}]^{1/2},
\end{eqnarray}
and $G_{\vec k}$ and $G_{\vec k}^{\pm}$ are defined as
\begin{eqnarray}
\label{eqB-1-6}
G_{\vec k}&=&E_{1}^{2}-E_{2}^{2}+F_{1}^{2}-F_{2}^{2}-2E_{3}^{2}+2E_{4}^{2},\nonumber\\
G_{\vec k}^{\pm}&=&\left(E_{3}^{2}-E_{4}^{2}\right)^{2}
+\left(E_{1}^{2}-E_{2}^{2}\right)\left(F_{1}^{2}-F_{2}^{2}\right)\nonumber\\
&&-2\left(E_{1}F_{1}-E_{2}F_{2}\right)(E_{3}^{2}+E_{4}^{2})\nonumber\\
&&\mp 4\left(E_{1}F_{2}-E_{2}F_{1}\right)E_{3}E_{4}.
\end{eqnarray}
The physical dispersion $\omega(\vec{k})$ is given by
either $4[\lambda_{1}(\vec k)+\lambda_{2}(\vec k)]$ 
or $4[\lambda_{3}(\vec k)+\lambda_{4}(\vec k))]$, depending on the
value of $\vec{k}$.
In the vicinity of the point $\vec k=\vec k_{h}$ ($\vec k_{H}$), 
which is the gapless points when $h = 0$ ($H = 0$), we find
\begin{eqnarray}
\label{eqB-1-7} 
\omega(\vec k)=4[\lambda_{1}(\vec k)+\lambda_{2}(\vec k)]. 
\end{eqnarray}
If only one of the external fields 
($H$ or $h$) is non-vanishing, the result is considerably simplified
because we have $E_{3}=E_{4}=0$. 
The spin-wave Hamiltonian can be actually diagonalized by a simpler
two-mode BT which is the same type as we need in the non-competitive case 
in Appendix~\ref{App2-2}. The resulting magnon dispersion is
\begin{widetext}
\begin{eqnarray}
\label{eqB-1-8}
\omega(\vec k) &=& \Big[\Big( 2SJ(1-2\sin^{2}\theta_{\rm cl}\cos^{2}\phi_{\rm cl})
+2S(J_{\perp}+J_{\perp}')(2\cos^{2}\theta_{\rm cl}\cos^{2}\phi_{\rm cl}-1)
-2SJ\sin^{2}\theta_{\rm cl}\cos^{2}\phi_{\rm cl}\cos k_{x}a_{x}\nonumber\\
&&+2S(\cos^{2}\theta_{\rm cl}(1+\sin^{2}\phi_{\rm cl})-1)
(J_{\perp}\cos k_{y}a_{y}+J_{\perp}'\cos k_{z}a_{z})
+ \Sigma 
\Big)^2\nonumber\\
&&-\Big(2SJ(\cos^{2}\theta_{\rm cl}-\sin^{2}\theta_{\rm cl}\sin^{2}\phi_{\rm cl})\cos k_{x}a_{x}
+2S\cos^{2}\theta_{\rm cl}\cos^{2}\phi_{\rm cl}
(J_{\perp}\cos k_{y}a_{y}+J_{\perp}'\cos k_{z}a_{z})\Big)^2
\Big]^{1/2},
\end{eqnarray} 
\end{widetext}
where $\Sigma = H\sin\theta_{\rm cl}\cos\phi_{\rm cl}$ when $h=0$,
and $\Sigma = h\sin\phi_{\rm cl}$ when $H=0$.
Of course, in these special cases,
Eq.~(\ref{eqB-1-7}) reduces to Eq.~(\ref{eqB-1-8}) near $\vec k =\vec k_{h}$
or $\vec k_{H}$.

From the dispersion (\ref{eqB-1-7}), 
let us estimate how $\Delta_{h}$ ($\Delta_{H}$) grows when a small
$h$ ($H$) is applied. 
To estimate $\Delta_{h}$, it is sufficient to know the coefficients  
of  Taylor expansion of $E_{1,2,3,4}(\vec k=\vec 0)$, $G_{\vec k=\vec 0}$ and 
$G_{\vec k=\vec 0}^{\pm}$ around $h=0$. As a result, in the limit of
small $h$, the gap behaves as 
\begin{eqnarray}
\label{eqB-1-10}
\Delta_{h} &\approx & 2S(J+J_{\perp}+J_{\perp}')^{1/2}(J_{\perp}+J_{\perp}')^{1/2}
(1-\tilde{H}^{2})^{1/2}\nonumber\\
&&\times\left(1-\tilde{H}^{2}+
\begin{array}{l}
\frac{J_{\perp}+J_{\perp}'}{J}
\end{array}
\tilde{H}^{2}\right)^{-1/2}\nonumber\\
&& \times\Big[\left(1-4\tilde{H}^{2}-
\begin{array}{l}
\frac{(J-J_{\perp}-J_{\perp}')^{2}}{J(J_{\perp}+J_{\perp}')}
\end{array}
\tilde{H}^{4}\right)^{1/2}\nonumber\\
&&+\left(1-
\begin{array}{l}
\frac{(J-J_{\perp}-J_{\perp}')^{2}}{J(J_{\perp}+J_{\perp}')}
\end{array}
\tilde{H}^{4}\right)^{1/2}\Big]\tilde{h}+\cdots, 
\end{eqnarray}
where $\tilde{H}=\frac{H}{4S(J+J_{\perp}+J_{\perp}')}$ 
and $\tilde{h}=\frac{h}{4S(J_{\perp}+J_{\perp}')}$.
At $H=0$, Eq.~(\ref{eqB-1-10}) is reduced to the exact result
$\Delta_{h}=[1+J/(J_{\perp}+J_{\perp}')]^{1/2}h$ which is 
derived from Eq.~(\ref{eqB-1-8}) and is drawn in Fig.~\ref{Fig9}.
Similarly to $\Delta_{h}$, $\Delta_{H}$ can be estimated. The result is   
\begin{eqnarray}
\label{eqB-1-11}
\Delta_{H} & \approx &2S(J+(J_{\perp}+J_{\perp}')(1-\tilde{h}^{2}))^{1/2}\nonumber\\
&&\times (J_{\perp}+J_{\perp}')^{1/2}(1-\tilde{h}^{2})^{-1/2}\nonumber\\
&&\times \Big[\left(1-
\begin{array}{l}
\frac{(J-J_{\perp}-J_{\perp}')^{2}}
{(J+(J_{\perp}+J_{\perp}')\tilde{h}^{2})(J+J_{\perp}+J_{\perp}')}
\end{array}
\tilde{h}^{2}\right)^{1/2}\nonumber\\
&&+\left(1-
\begin{array}{l}
\frac{J+J_{\perp}+J_{\perp}'}{(J+(J_{\perp}+J_{\perp}')\tilde{h}^{2})}
\end{array}
\tilde{h}^{2}\right)^{1/2}
\Big]\nonumber\\
&&\times\tilde{H}+\cdots,
\end{eqnarray}
which is reduced to the conventional result $\Delta_{H}=H$ when $h=0$.
Both gaps have linear field dependence.
The results (\ref{eqB-1-10}) and (\ref{eqB-1-11}) indicate that
$\Delta_{h}$ ($\Delta_{H}$) is non-vanishing only when 
$h\neq 0$ ($H\neq 0$). 
As a consequence, the true gap $\Delta={\rm min}(\Delta_h, \Delta_H)$ is
zero when either of $h$ or $H$ is zero.

\subsection{The non-competitive case}
\label{App2-2}
Here we summarize the spin-wave approximation on the non-competitive case.
We find that it can be straightforwardly applied also to the 
symmetric phase in the competitive case.
   
According to the MFT and Fig.~\ref{Fig7}, the spin configuration
in the classical ground state can be written as
\begin{eqnarray}
\vec{S}_{2n,j,k} &\sim& S R_{y}(-\theta)  \hat{x}, \\
\vec S_{2n+1,j,k} &\sim& S R_{z}(\pi)R_{y}(-\theta) \hat{x},
\end{eqnarray}
where $\hat{x}$ is the unit vector pointing to $x$ direction
and $\theta$ is the canting angle parameter.
The minimization of the classical energy determines
$\theta=\theta_{\rm cl}$ as
\begin{equation}
\label{eqB-2-3}
\cos\theta_{\rm cl}\sin \theta_{\rm cl}=\frac{H}{4SJ}\cos\theta_{\rm cl}
-\frac{h}{4SJ}\sin\theta_{\rm cl}.
\end{equation}

A standard spin-wave theory on this classical ground state gives
the quadratic Hamiltonian in terms of bosons
\begin{eqnarray}
\label{eqB-2-4}
\hat H_{\rm HP2}&=&\sum_{\vec k}\left(
\begin{array}{ll}
a_{\vec k}^{\dag}& a_{-\vec k}
\end{array}
\right)
\left(
\begin{array}{cc}
\frac{1}{2} A_{\vec k} & B_{k_{x}}\\
B_{k_{x}}& \frac{1}{2}A_{\vec k}
\end{array}
\right)
\left(
\begin{array}{l}
a_{\vec k}\\
a_{-\vec k}^{\dag}
\end{array}
\right)\nonumber\\
&&-\frac{1}{2} A_{\vec k}
\,\,\,+\mbox{const},
\end{eqnarray}
where
\begin{eqnarray}
\label{eqB-2-5}
\begin{array}{lll}
A_{\vec k}=A_{-\vec k}&=&2SJ \cos 2\theta_{\rm cl}-2SJ\sin^{2}\theta_{\rm cl}\cos k_{x}a_{x}\\ 
&&+2S|J_{\perp}|(1-\cos k_{y}a_{y})\\
&&+2S|J_{\perp}'|(1-\cos k_{z}a_{z})\\
&&+H \sin\theta_{\rm cl}+h \cos\theta_{\rm cl},\\
B_{k_{x}}=B_{-k_{x}}&=& -SJ\cos^{2}\theta_{\rm cl}\cos k_{x}a_{x}.
\end{array}
\end{eqnarray}
Now we apply the two-mode BT
\begin{equation}
\label{eqB-2-6}
\left(
\begin{array}{cc}
a_{\vec k} \\
a_{-\vec k}^{\dag}
\end{array}
\right)=
\left(
\begin{array}{cc}
u_{\vec k} & v_{\vec k} \\
v_{\vec k} & u_{\vec k}
\end{array}
\right)
\left(
\begin{array}{cc}
\tilde{a}_{\vec k} \\
\tilde{a}_{-\vec k}^{\dag}
\end{array}
\right),
\end{equation}
where $u_{\vec k}^{2}-v_{\vec k}^{2}=1$, to Eq.~(\ref{eqB-2-4}).
The Hamiltonian is then diagonalized as
\begin{equation}
\label{eqB-2-7}
\hat H_{\rm HP2} = \sum_{\vec k}\tilde{\omega}(\vec k)\,
\tilde{a}_{\vec k}^{\dag}\,\tilde{a}_{\vec k}+{\rm const.}~,
\end{equation}
if we choose $u_{\vec k}^{2} = \frac{1}{2}
(1+\frac{A_{\vec k}}{\tilde{\omega}(\vec k)})$ and $v_{\vec k}^{2}= 
\frac{1}{2}(-1+\frac{A_{\vec k}}{\tilde{\omega}(\vec k)})$.
The dispersion relation is given by
\begin{equation}
\label{eqB-2-8}
\tilde{\omega}(\vec k)=\sqrt{A_{\vec k}^{2}-4 B_{k_{x}}^{2}}.
\end{equation}
The above derivation and results apply exactly to the 
symmetric phase in the competitive case, only with the
replacement
$(|J_{\perp}|,|J_{\perp}'|)\rightarrow (-J_{\perp},-J_{\perp}')$.
The dispersion (\ref{eqB-2-8}) has the gapless point $\tilde{\omega}(0,0,0)$ 
in the AF phase where $h=0$. It corresponds to the NG mode
due to the spontaneous breaking of the U(1) symmetry.
Hence $\tilde{\Delta}_{h}= \tilde{\omega}(0,0,0)$ can be regarded as
the $h$-induced gap.
At the small field $h$, the angle variation around $h=0$: 
$\delta \theta \equiv \sin^{-1}(\frac{H}{4JS})-\theta_{\rm cl}$ 
is estimated approximately as  
\begin{equation}
\label{eqB-2-11}
\delta \theta \approx -\frac{H}{(4SJ)^{2}-H^{2}}\,\,h.
\end{equation}
Therefore expanding $\tilde{\Delta}_{h}^{2}(h,\delta \theta)$ around $h=\delta \theta=0$, 
one sees that the gap grows as
\begin{equation}
\label{eqB-2-12}
\tilde{\Delta}_{h} \approx \sqrt{4SJ}\left[1-\left(\frac{H}{4SJ}\right)^{2}\right]^{1/4}
\left[1+2\left(\frac{H}{4SJ}\right)^{2}\right]^{1/2}h^{1/2}.
\end{equation}
In contrast to the SSB phase (in the competitive case),
the $h$-induced gap opens as $\tilde{\Delta}_{h} \propto h^{1/2}$. 
Furthermore, it is remarkable that Eq.~(\ref{eqB-2-12}) has no dependence on
the inter-chain interactions.
In fact, it is identical to the 1D result [Eq.~(3.17) in
Ref.~\onlinecite{A-O}]. This is a reflection of the smoothness which was
discussed in the final of Sec.~\ref{Sec3}. 
At $H=0$, we can obtain the simple exact result 
$\tilde{\Delta}_{h}=\sqrt{4SJh}[1+h/(4SJ)]^{1/2}$ from the dispersion (\ref{eqB-2-8}).

\end{document}